\documentclass[11pt]{article}
\pdfoutput=1 
\usepackage{amssymb}
\usepackage[margin=1in]{geometry}

\usepackage{lineno}
\usepackage{graphicx}

\usepackage{moreverb,url}
\usepackage{makecell}
\usepackage[breaklinks=true, colorlinks=true, linkcolor={blue!90!black}, citecolor={blue!90!black}, urlcolor={blue!90!black}]{hyperref}

\newcommand\BibTeX{{\rmfamily B\kern-.05em \textsc{i\kern-.025em b}\kern-.08em
T\kern-.1667em\lower.7ex\hbox{E}\kern-.125emX}}

\usepackage[algo2e,linesnumbered,vlined,algoruled]{algorithm2e}

\usepackage{placeins}
\SetCommentSty{mycommfont}

\usepackage{geometry,xcolor,tabularx,nccmath}
\usepackage{amsmath,amsthm,amssymb,enumitem}
\usepackage{booktabs,caption,subcaption,mathtools,multirow}
\usepackage{graphicx,tabulary}
\usepackage{rotating}

\usepackage[utf8]{inputenc}
\usepackage{amsfonts}
\usepackage{fullpage}

\newcommand{\f}{f_{u^*,v^*,i^*,j^*}}

\usepackage[colorinlistoftodos]{todonotes}

\usepackage{placeins}
\usepackage{etoolbox}

\usepackage[nameinlink]{cleveref}
\hypersetup{colorlinks={true},allcolors={blue}}

\crefformat{section}{\S#2#1#3}
\crefformat{subsection}{\S#2#1#3}
\crefformat{equation}{#2Equation~\textup{(#1)}#3}
\crefformat{cons}{#2\textbf{Constraint}~\textup{\textbf{(#1)}}#3}
\labelcrefformat{cons}{\textup{#2(#1)#3}}
\crefformat{consset}{#2\textbf{Constraints}~\textup{\textbf{(#1)}}#3} 
\labelcrefformat{consset}{\textup{\textbf{#2(#1)#3}}}
\crefformat{func}{#2Function~\textup{(#1)}#3}
\labelcrefformat{func}{#2Function~\textup{(#1)}#3}
\crefformat{algo}{#2\textbf{Algorithm}~\textup{\textbf{#1}}#3}
\crefformat{objfunc}{#2\textbf{Objective function}~\textup{\textbf{(#1)}}#3}
\labelcrefformat{objfunc}{\textup{(#2#1#3)}}
\crefformat{ch}{#2Chapter~\textup{#1}#3}
\crefformat{sec}{#2Section~\textup{#1}#3}
\crefformat{tab}{#2Table~\textup{#1}#3}
\crefformat{fig}{#2Figure~\textup{#1}#3}
\crefformat{eq}{#2equation~\textup{(#1)}#3}
\labelcrefformat{eq}{\textup{(#2#1#3)}}
\crefformat{theo}{#2Theorem~\textup{#1}#3}
\crefformat{lem}{#2Lemma~\textup{#1}#3}
\crefformat{app}{#2Appendix~\textup{#1}#3}
\crefformat{scenario}{#2Scenario~\textup{#1}#3}
 \AtBeginDocument{%
    \crefname{figure}{Figure}{figures}%
}

\expandafter\def\expandafter\UrlBreaks\expandafter{\UrlBreaks
  \do\a\do\b\do\c\do\d\do\e\do\f\do\g\do\h\do\i\do\j%
  \do\k\do\l\do\m\do\n\do\o\do\p\do\q\do\r\do\s\do\t%
  \do\u\do\v\do\w\do\x\do\y\do\z\do\A\do\B\do\C\do\D%
  \do\E\do\F\do\G\do\H\do\I\do\J\do\K\do\L\do\M\do\N%
  \do\O\do\P\do\Q\do\R\do\S\do\T\do\U\do\V\do\W\do\X%
  \do\Y\do\Z}

\usepackage{amsmath,cases,eqparbox,xparse}

\makeatletter
\NewDocumentCommand{\eqmathbox}{o O{c} m}{%
  \IfValueTF{#1}
    {\def\eqmathbox@##1##2{\eqmakebox[#1][#2]{$##1##2$}}}
    {\def\eqmathbox@##1##2{\eqmakebox{$##1##2$}}}
  \mathpalette\eqmathbox@{#3}
}
\makeatother

\makeatletter
\let\@msm@th@eqref\eqref
\renewcommand{\eqref}[1]{%
  \begingroup
  \leavevmode
  \color{blue}%
  \hypersetup{linkbordercolor=[named]{blue}}%
  \@msm@th@eqref{#1}%
  \endgroup
}
\makeatother

\usepackage[normalem]{ulem} 
 
\crefname{defi}{definition}{definitions}
\Crefname{defi}{Definition}{Definitions}
\crefname{lemma}{lemma}{lemmas}
\Crefname{lemma}{Lemma}{Lemmas}
\crefname{assumption}{assumption}{assumptions}
\Crefname{assumption}{Assumption}{Assumptions}

\usepackage{multicol}

\usepackage{authblk}
\providecommand{\keywords}[1]{\noindent\textbf{Keywords:} #1}

\usepackage{natbib}

\begin{document}

\title{Modeling Transit in a Fully Integrated Agent-Based Framework: Methodology and Large-Scale Application}


\author[]{Omer Verbas*}
\author[]{Taner Cokyasar}
\author[]{Pedro Veiga de Camargo}
\author[]{Krishna Murthy Gurumurthy}
\author[]{Natalia Zuniga-Garcia}
\author[]{Joshua Auld}
\affil[]{Argonne National Laboratory, 9700 S. Cass Ave., Lemont, IL 60439, USA }
\affil[]{*Corresponding Author}

\maketitle


\begin{abstract}
This study presents a transit routing, assignment, and simulation framework which is fully embedded in a multimodal, multi-agent transportation demand and supply modeling platform. POLARIS, a high-performance agent-based simulation platform, efficiently integrates advanced travel and freight demand modeling, dynamic traffic and transit assignment, and multimodal transportation simulation within a unified framework. We focus on POLARIS's transit routing, assignment, and simulation components, detailing its structural design and essential terminologies. We demonstrate how the model integrates upstream decision-making processes — activity generation, location and timing choices, and mode selection, particularly for transit-inclusive trips — followed by routing, assignment decisions, and the movement of travelers and vehicles within a multimodal network. This integration enables modeling of interactions among all agents, including travelers, vehicles, and transportation service providers. The study reviews literature on transportation system modeling tools, describes the transit modeling framework within POLARIS, and presents findings from large-scale analyses of various policy interventions. Results from numerical experiments reveal that measures such as congestion pricing, transit service improvements, first-mile-last-mile subsidies, increased e-commerce deliveries, and vehicle electrification significantly impact transit ridership, with some interactions between these levers exhibiting synergistic or canceling effects. The case study underscores the necessity of integrating transit modeling within a broader multimodal network simulation and decision-making context.
\end{abstract}


\section{Introduction} \label[sec]{intro}
Public transit plays a crucial role in modern transportation systems, serving as a core component of urban mobility and substantially contributing to the overall efficiency of travel networks. In addition to moving people from one point to another and reducing traffic congestion, transit services, such as buses, trains, subways, and trams, bring in various benefits to economy, environment, social justice. Transit lowers greenhouse gas emissions and provides accessibility for diverse populations. By integrating transit into urban environments, we do not only enhance mobility but also promote sustainable development and improve the quality of life for all residents. As urban areas grow, maintaining and enhancing transit services while integrating them with other transportation modes is key to control traffic congestion and to provide an equitable access to mobility for all.

Due to its value in urban life, transit has to be a key component of any transportation simulation analysis tool. In POLARIS, (Planning and Operations Language for Agent-based Regional Integrated Simulation) \cite{auld2016polaris}, transit network is simulated in tandem with other travel modes, e.g., automobile, bike, truck, and walk. POLARIS is a high-performance and computationally efficient agent-based simulation tool that implements advanced travel and freight demand modeling, dynamic traffic and transit assignment, and transportation simulation in an integrated modeling platform \cite{POLARIS_website}. It is used to analyze different transportation system management strategies including emerging vehicle and information technologies and to investigate the consequent impact on economy, energy, equity, and mobility utilizing a seamlessly integrated framework with sibling analysis tools, e.g., Autonomie, RoadRunner, and SVTrip, developed at Argonne National Laboratory, a federally funded research and development center supported by the U.S. Department of Energy \cite{VMS_tools_website}.

In this study, we focus on transit routing, assignment, and simulation components of POLARIS and present its structural design while describing key terminologies and utilized concepts. The upstream decision-making of travelers, such as activity generation, location and timing choices, and mode choice (here, we focus on trips where at least a portion of the journey is through transit) are captured in a fully integrated fashion. These are followed by the downstream decision-making of routing and assignment, followed by the movement of travelers and vehicles in the multimodal network. This integrated framework enables us to model the various interactions between all agents (freight, cars, travelers, transit agencies etc.) in its full extent.

The rest of the paper is organized as follows. After a brief literature review, we present the overall POLARIS framework, its details on transit and active network layers, and the routing, assignment, and simulation modules with a focus on transit, in the methodology section. The numerical experiments are transit-related details from a large-scale study in \citep{auld2024large}. The last section concludes the paper.

\section{Literature Review} \label[sec]{lit_rev}

Simulation in public transportation enables a thorough examination of the complex relationships among various elements of the system, such as advanced transportation technologies, infrastructure, passengers, and vehicles. While there are numerous existing simulation tools—both commercial and open-source—that can assess and refine control systems, creating platforms that address the unique needs of public transit remains a challenging and ongoing area of research and development.

\citep{algers1996review} and \citep{boxill2000evaluation} reviewed traffic simulation models and software at large. Their studies provided a list of software used by transportation planning agencies and researchers. Yet, the list is not up-to-date, and some of the mentioned tools do not exist or are no longer maintained. Much more recently, \citep{lovelace_2021} provided a comprehensive review of open-source tools for transport planning, although many of those have evolved substantially since the article's publication.

In \cite{ghariani2014survey}, the authors provided a comprehensive review of simulation tools utilized for evaluating public transit control systems. The authors focused on how these tools aid in assessing the efficiency of various public transportation strategies and operations. Moreover, they referenced a list of commonly used transit simulation tools to the date of publication.

PTV Visum is a commercial macroscopic transport modelling software for traffic planning, offering a comprehensive and detailed depiction of all public transportation modes, including buses, trams, subways, taxis, railways, and trains \cite{PTV_visum}. The tool is commonly used by transportation planners and analysts all over the world \cite{heyken2021public}. It can create cost-effective and demand-oriented transport service designs while creating new routes or stations and optimizing service schedules. 

Paramics Microsimulation is a 3D simulation tool that enables testing new infrastructure improvements, road network operation, traffic signal control optimization, public transport operations, and economic and environmental assessment \cite{Paramics}.

Aimsun (Advanced interactive microscopic simulator for urban and non-urban networks) commercial simulation tool provides various platforms to address transportation questions from a data-driven and simulation perspective \cite{Aimsun}. Aimsun Insight identifies patterns and transforms big data into mobility insights. Aimsun Predict forecasts future traffic conditions powered by real-time data. Aimsun Start uses open-source data to simulate transportation scenarios at high-fidelity. Aimsun Plus is an extended scale of Aimsun Start, covering a city or a regional transportation simulation and is similar to POLARIS from a scale perspective. Aimsun Live fuses the other four platforms in a single one and simulates an entire large-scale mobility network utilizing real-time data.

TransCAD transportation planning software is a commercial Geographic Information System (GIS) tool that offers a platform for managing, analyzing, and visualizing transportation data \cite{TransCAD}. It contains tools for mapping, visualization, and analysis, along with application modules for routing, travel demand forecasting, public transit, logistics, site location, and territory management. 

Aequilibrae \cite{aequilibrae_2023} is an open-source modelling package with similar capabilities to many of the packages above though more focused on traditional techniques, such as static assignment, route choice, and frequency-based transit assignment, rather than micro-simulation.

Eclipse SUMO (Simulation of Urban MObility) is an open-source microscopic and continuous multimodal traffic simulation tool capable of handling large networks \cite{Sumo}. The tool is adopted in many research studies, e.g., \cite{khumara2018estimation,krajzewicz2012recent,monga2022sumo}.

MATSim is an open-source platform capable of handling large networks \cite{matsim_2016} that has been adopted in multiple studies \cite{bhatmatsim, NOVOSEL, miller_matsim}. Differently than SUMO, however, MATSim has a bigger focus on demand and other transport modes such as transit and Transportation Network Companies (TNC).

Behavior, Energy, Autonomy, Mobility Comprehensive Regional Evaluator (BEAM CORE) is an open-source agent-based regional transportation model developed by Lawrence Berkeley National Laboratory and its collaborators to analyze complex transportation systems and emerging mobility trends \cite{spurlock2024behavior}. The tool enables multi-year scenario analyses by integrating various modules that capture changes in land use, population dynamics, vehicle technology adoption, and freight logistics.

Northwestern University Transit Assignment and Simulation (NUTRANS) \cite{verbas2014dissertation, verbas2015hyperpath, verbas2015trr_dta, verbas2015integrated_freq_dta, verbas2016partb_dta} is a Simulation-Based Dynamic Transit Assignment tool. It utilizes a hyperpath algorithm for path-finding, a gap-based approach for assignment, and a multi-agent particle simulation module for the movement of travelers and transit vehicles. The model has been integrated with DYNASMART \cite{jayakrishnan1994evaluation} for co-simulation of roadway and transit networks in an integrated framework.

There are two more sophisticated components that are usually overlooked in these transit modelling frameworks, however. The first one is the co-simulation of public transport and traffic, often replaced by a feedback mechanism that tries to make both components, transit and traffic, consistent. The second component is what can be broadly called multi-modal routing (and assignment). This issue has received substantial attention in recent years, particularly due to the rise of micro-mobility and TNCs. Still, it is yet to be determined whether more simplified simulated frameworks, as well as non-simulated frameworks, will be capable of dealing with this issue in a consistent manner.

\section{Methodology} \label[sec]{methodology}

POLARIS \cite{auld2016polaris} is a software platform that combines activity-based travel demand modeling with multimodal traffic and transit assignment in an agent-based framework. It simulates the movements of individuals and freight over a 24-hour period. This section presents a brief summary of overall POLARIS workflow, followed by the details on multimodal network representation and its generation. The section concludes with descriptions of multimodal routing, assignment, and simulation.

\subsection{POLARIS Workflow}\label[sec]{POLARIS_overview}

\begin{figure}[!h]
    \centering
    \includegraphics[width=1\linewidth]{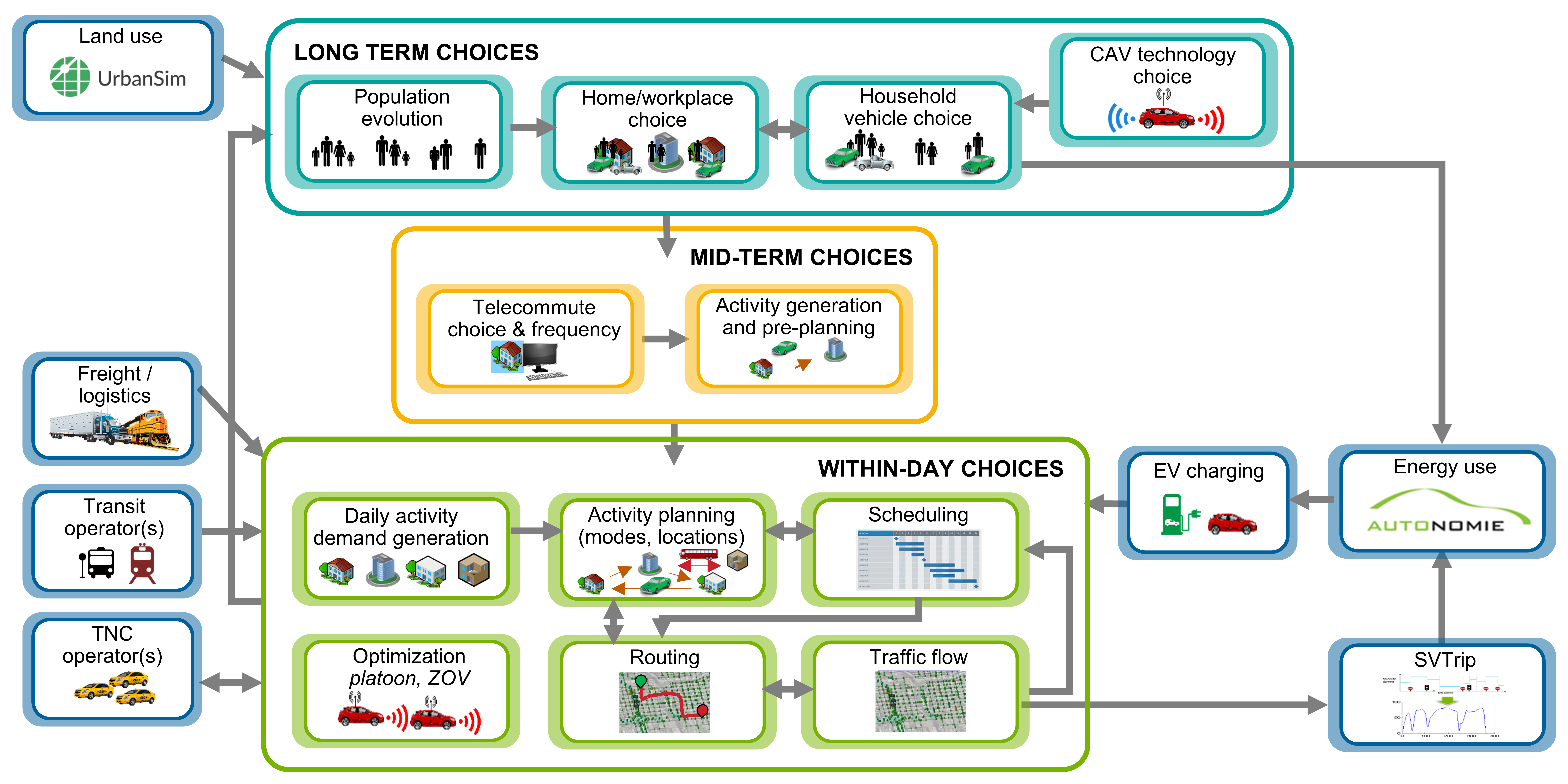}
    \caption{POLARIS workflow.}
    \label[fig]{POLARIS_Workflow}
\end{figure}

This subsection summarizes the overall POLARIS workflow. See \cref{POLARIS_Workflow} for details on POLARIS workflow.

\subsubsection{Activity-Based Demand Modeling}\label[sec]{ABM_demand}
POLARIS creates a synthesized population of households and individuals using data from U.S. Census tracts, Public Use Microdata Areas (PUMAs), and the American Community Survey (ACS). This process involves selecting home, school, and work locations \cite{31auld2010efficient}. Activities for each individual are generated for a 24-hour period based on socioeconomic details at both personal and household levels. The start time and duration of these activities are determined using a hazard-based model \cite{32auld2011dynamic}, while a multinomial logit model is used to choose activity locations \cite{33auld2012activity}, and a nested logit model is used to select travel modes \cite{33auld2012activity}. Additionally, the model accounts for the choice of travel party \cite{34auld2011planning} . A conflict resolution model is employed to address scheduling conflicts within an individual's plans and among household members \cite{35auld2009modeling}.

\subsubsection{Freight Modeling}\label[sec]{freight}
Similar to passenger travel, POLARIS synthesizes firms and establishments using the CRISTAL model \cite{36stinson2022introducing}. These firms and establishments generate truck and e-commerce delivery trips \cite{38cokyasar2023time,39zuniga2023freight}. Although the specifics of these models are not covered in this study, it is crucial to highlight that transportation operates as a complex-adaptive system where each agent's decisions influence the decisions of others. For instance, a decrease in the number of delivery trucks would improve passenger car traffic causing a modal shift to passenger cars, and thus negatively impact transit ridership.

\subsubsection{Multimodal Supply Modeling}\label[sec]{supply}
Once the population is synthesized and individual activity schedules, including timing, location, mode, and travel party, are generated, and all truck and delivery trips are created using the CRISTAL module, the simulation begins. Each trip is routed using a time-dependent intermodal A* algorithm \cite{verbas2018polaris_routing}. The network equilibrium is achieved within a dynamic transit and traffic assignment framework by integrating information from both current and historical network conditions \cite{auld2019agent}. The details of the routing and assignment will be discussed in more detail in the following subsections.

Roadway traffic is simulated with a multi-class traffic flow model in Lagrangian coordinates \cite{deSouza2024polarisLC}, where each vehicle class has a distinct speed-spacing relationship. This extends the mesoscopic traffic flow model \cite{deSouza2019mesoscopic} previously used in POLARIS, which employed a single average speed-spacing relationship per link based on the traffic mix. Transit buses operate on the congestable roadway network, while trains run on a non-congestable network following the General Transit Feed Specification (GTFS) \cite{GTFS_reference} travel times.

POLARIS also simulates the movements and decision-making processes of TNC agents for various services, including single and pooled rides, corner-to-corner trips, first-mile/last-mile (FMLM) transit integration, on-demand transit services, and more centralized and autonomous business models that could emerge in the future \cite{gurumurthy2020integrating,gurumurthy2021system,zuniga2022integrating}.

\subsection{Multimodal Network Representation and Generation}\label[sec]{network_rep}

\begin{sidewaysfigure}[!htbp]
    \centering
    \includegraphics[width=1\textwidth,height=1\textheight,keepaspectratio]{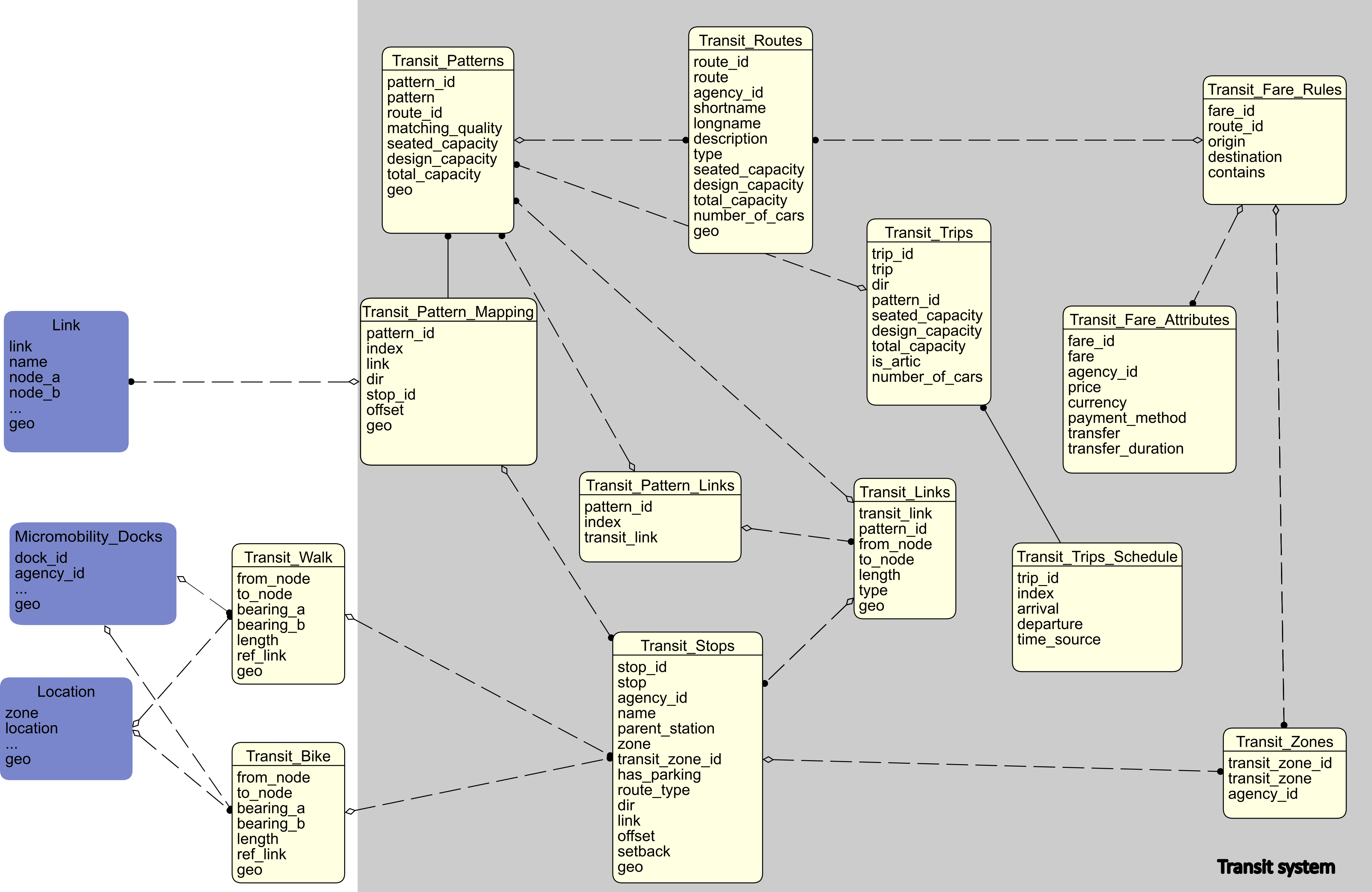}
    \caption{Transit database schema in POLARIS.}
    \label[fig]{db_schema}
\end{sidewaysfigure}

This subsection focuses on adding transit and active mode representations to an existing POLARIS network with roadways only. Transit models in POLARIS are generally created by importing GTFS \cite{GTFS_reference} feeds generated by planning agencies or transit operators, as that is both the most accurate representation of real-world transit services and the most convenient form to ingest the data. The GTFS data mainly consists of stops/stations, routes, trips, and stop times, i.e. schedules. The data is not imported as-is, and a series of transformations and consistency checks are executed as the data is prepared to be inserted into the model. See \cref{db_schema} for an overview of all transit and related tables and how they connect to each other through foreign key references.

There are two drivers behind the change in the data structure during the data import.  The first one is the need to have the data in a format convenient for computation, which in this case means generating a network graph that can be quickly loaded into memory. The stop/station data is converted to \textit{Transit Stops} table almost one-to-one. These are to become the \textit{transit nodes} in the multimodal network representation. To generate links, the schedule file is utilized. Any two consecutive stops for a given trip form a link. 

\begin{figure}[!h]
    \centering
    \includegraphics[width=.5\linewidth]{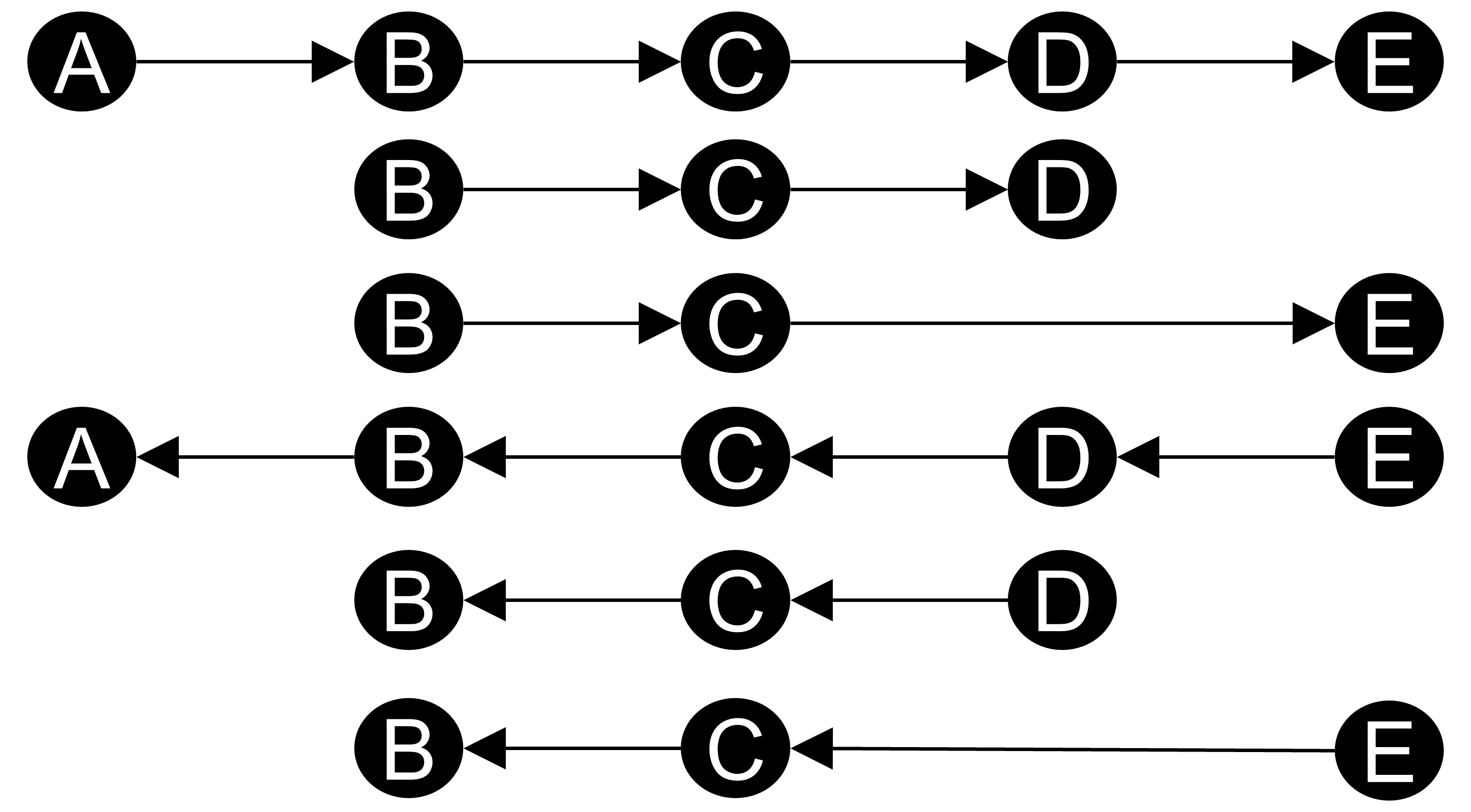}
    \caption{A sample of six different transit patterns for a given route.}
    \label[fig]{patterns}
\end{figure}

The second key reason for data conversion is the creation of \textit{Transit Patterns}. Transit routes can be interpreted as a sequence of transit stops or transit links. However, many large-scale agencies utilize the concept of \textit{Transit Patterns}, which is a unique sequence of stops that is served by a certain transit route. As seen in \cref{patterns}, a transit route may serve between stops A and E in either direction. The first and fourth pattern in \cref{patterns} are \textit{full-route} patterns, the second and fifth are \textit{short-turning} patterns \cite{furth_short,ceder1989optimal}, whereas the third and sixth are \textit{express service} \cite{furth_zonal} patterns. They are utilized by transit agencies for more efficient operations to focus on higher demand origin-destination stop pairs. Many studies in the literature have acknowledged and incorporated the concept of service patterns into their study framework \cite{verbas2013trr_freq, verbas2014dissertation, verbas2015stretching, verbas2015partb_freq, verbas2015hyperpath, verbas2015trr_dta, verbas2015integrated_freq_dta, verbas2016partb_dta, verbas2016mode_choice, verbas2018polaris_routing, auld2019agent, pinto2020joint}.

While some agencies may explicitly declare their service patterns in the GTFS using \textit{shape IDs}, many agencies do not. As a result, the procedure parses through every trip's stop sequence, and for every route, it detects the unique sequence of stops, which generates a set of unique patterns for every route. Each route is a unique row entry in the \textit{Transit Routes} table, whereas each pattern is a unique row entry in the \textit{Transit Patterns} table (See \cref{db_schema}). \textit{Transit Pattern Links} table lists the link sequence of every pattern, whereas \textit{Transit Links} table carries additional information on each link. Please note that each [from node, to node, pattern ID] tuple is a unique link in this network representation. \textit{Transit Patterns} can be interpreted as \textit{precisely defined routes}, and its importance in the POLARIS data model is clear from the schema on \cref{db_schema}.

Utilizing the calendar and calendar dates files, and \textit{service\_id}s, one can find which trips are active on a given date, which can be a typical weekday. Those trips form single-row entries in the \textit{Transit Trips} table, and their schedule is listed in the \textit{Transit Trips Schedule} table. If a given transit pattern has \textit{k} links, hence \textit{k} rows in the \textit{Transit Pattern Links} table, then it has \textit{k+1} rows in the \textit{Transit Trips Schedule} table, because the latter has as many entries as the number of stops.

Regarding checks, the first verification conducted during the import process is an internal consistency check similar to that done by the canonical GTFS validator maintained by the Mobility Data organization \cite{mobilitydata} and other general-purpose GTFS validators. In the case of the POLARIS, these checks are related to potential errors that would cause the POLARIS to malfunction, such as the existence of trips for routes that do not exist and route types not contemplated in the standard.

The most relevant data check, however, consists of verifying vehicle speeds between consecutive stops. During this verification step, the speed of each transit segment is compared to the maximum speed allowed to that particular transit mode and segment length (shorter segments have lower average speeds due to acceleration/deceleration), and stop arrivals at all subsequent stops are delayed/updated to reduce average speed to the maximum allowed when it exceeds it. It is noteworthy that these maximum values are context-specific and, therefore, should be derived from transit data for each newly modeled city. The POLARIS online documentation \cite{polaris_pt_docs} contains the derived maximum speed values for Atlanta, Austin, and Chicago and include buses, heavy rail, and metro. There are also various checks after the data is fully imported such as link and stop count consistencies in the \textit{Transit Pattern Links} and \textit{Transit Trips Schedule} tables, respectively; and other types of consistencies among tables.

The POLARIS transit data model also includes fare information, which is almost a direct import from the GTFS database. Interested readers are encouraged to study \citep{GTFS_reference} for further details.

Another important aspect is the existence of multiple agencies in the study area. The importer seamlessly combines these entries by generating a \textit{Transit Agencies} table. A reference to this table is omitted in \cref{db_schema} to reduce illustration complexity. However, each agency receives a unique integer ID called \textit{agency ID}, and every route and stop are associated with their corresponding \textit{agency ID}. The rest of the affiliations flow through foreign key references. An example is that every trip is associated with a unique pattern, every pattern is associated with a unique route, and every route is associated with a unique agency. Please also note that \textit{x\_id} refers to a unique integer used in POLARIS identifying the primary key in most tables, whereas \textit{x} is the original ID shown in the GTFS database. As an example, \textit{route} in the \textit{Transit Routes} table refers to the route identifier published in the GTFS database, whereas \textit{route\_id} is a unique, integer, POLARIS identifier.

\begin{figure}[!h]
    \centering
    \includegraphics[width=.75\linewidth]{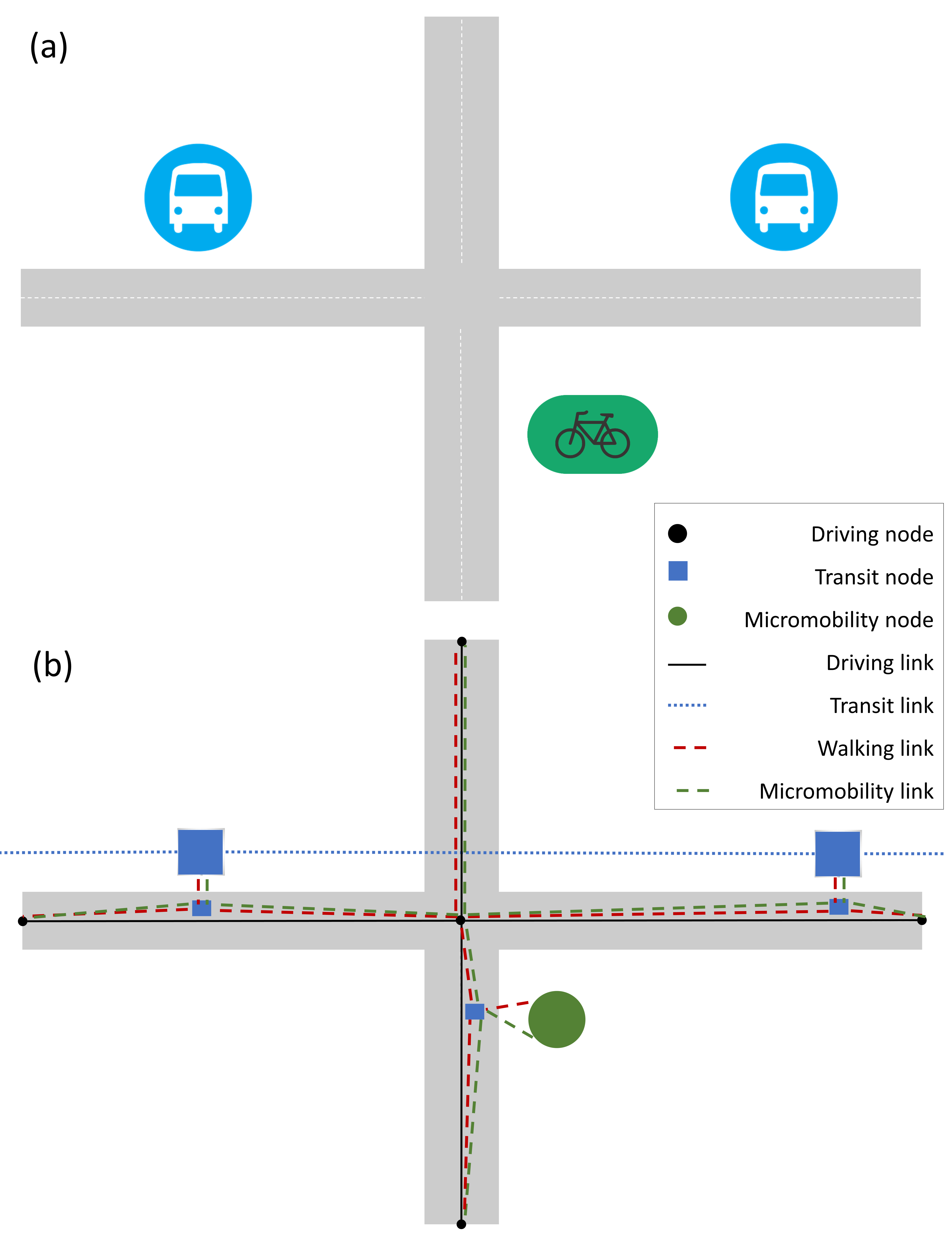}
    \caption{A sample multimodal network (a) in the real-world (b) as a multimodal graph in POLARIS.}
    \label[fig]{network}
\end{figure}

The POLARIS transit data model also includes separate tables for walk and bike (micromobility) access to transit. These two tables, shown outside the set of tables dedicated to the transit system data in \cref{db_schema}, are generated through post-processing. See \cref{network} for a simple example. There are four roadway segments with a four-way intersection, two bus stops in the East to West direction, and one bicycle dock in the South-North direction (\cref{network}a). As seen in \cref{network}b, there are five \textit{driving} (roadway) nodes and four \textit{driving} (roadway) links, which are represented by black dots and black lines, respectively. The blue squares represent the \textit{transit} nodes, whereas the blue lines represent the \textit{transit} links. Initially those two sub-networks, i.e., the \textit{driving} and \textit{transit} sub-networks are not connected. Keeping the nodes, the \textit{driving} links are copied once to generate the \textit{walking} links. Non-walkable links such as freeways, ramps etc. are removed. If there are additional walking links, those can be added through a separate procedure. At this stage, the \textit{walking} and \textit{driving} sub-networks are connected, where the \textit{driving} nodes are the commonly used nodes to facilitate transfer between sub-networks. Next, the \textit{transit} and \textit{micromobility} nodes are projected onto the \textit{walking} links. At the point of projection, new nodes are generated which bisect the link. Additional (virtual) \textit{walking} links are generated to connect the new node to the original \textit{transit} node. At this stage, the \textit{walking}, \textit{driving}, and \textit{transit} sub-networks are fully connected. The same procedure is repeated to generate the \textit{micromobility} sub-network and connect it with the rest of the network. With this final step, all four sub-networks are fully integrated.

Finally, POLARIS transit networks can be fully edited through a comprehensive Python API that allows:
\begin{itemize}
    \item Modifying existing transit pattern sequences using the existing stops,
    \item Defining new transit patterns for a given route using the existing stops,
    \item Adding new stops,
        \begin{itemize}
            \item Using those stops to create new routes,
                \begin{itemize}
                    \item In case of bus rapid transit (BRT) routes, dedicating lanes to BRT,
                \end{itemize}
            \item Using those stops to create new patterns for existing routes,
            \item Using those stops to modify existing patterns,            
        \end{itemize}
    \item Removing stops and then correctly updating all the pattern and trip schedule information,
    \item Updating frequencies,
        \begin{itemize}
            \item Manually or using certain rules applying to certain agencies, routes, time-windows etc.,
            \item Using optimization techniques \cite{verbas2013trr_freq,verbas2015integrated_freq_dta,verbas2015partb_freq}    
        \end{itemize}
    \item Updating speeds.
\end{itemize}

\subsection{Multimodal Routing, Assignment, and Simulation}\label[sec]{routing_assignment_sim}
This subsection summarizes the intermodal routing algorithm and assignment procedures implemented on a multimodal network representation in POLARIS. For details on the routing algorithm in POLARIS, the readers are encouraged to read \citep{verbas2018polaris_routing}. Similarly, the details of the assignment algorithm can be found in \citep{auld2019agent}.

\subsubsection{Time-Dependent Intermodal A* Algorithm}\label[sec]{routing_details}
Routing algorithms are classified in different ways. Label-setting algorithms \cite{dijkstra1959} update every node only once, whereas label correcting algorithms \cite{ford_algorithm, bellman_routing} allows for reevaluating a node  and updating its label. Initial algorithms are static, i.e., the network conditions do not change over time. The first time-dependent algorithm was introduced by \citep{cooke1966shortest}. Other important aspects are multi-modality and movement/transfer penalties. In the case of a roadway network, movement penalty is associated with making a turn. In the case of multi-modality, it can be associated with the cost of transferring between modes, or waiting for a transit vehicle, and so on. \citep{ziliaskopoulos2000intermodal} introduced a  multi-modal, time-dependent algorithm with movement/transfer penalties. These algorithms result in solutions where link-path incidence matrix has binary values. A link either belongs to a path or not. Relaxing the binary condition results in a probabilistic tree, where every link has a certain probability to be taken. These are called \textit{hyperpaths} \cite{nguyen1988equilibrium,nguyen2006hyperpaths} or \textit{optimal strategies} \cite{spiess1989optimal}. 

All of these algorithms mentioned up to this point are tree-based. The algorithm is either rooted at an origin and provides the shortest path from that origin to all destinations or vice-versa. \citep{hart1968formal} introduced the first point-to-point \textit{(A*)} algorithm, that finds an exact path between any two nodes in a static network. For time-dependent versions, interested readers can refer to \cite{chabini2002adaptations,zhao2008algorithm}. An A* algorithm for transit was introduced by \citep{bander1991new}. All A* path-finders utilize a heuristic-cost estimation process to guide the algorithm towards the destination. \citep{khani2015trip} introduced a \textit{transit-oriented} heuristic cost estimation process, where, instead of using a Euclidean-based estimator, they calculate the minimum possible transit travel time between the evaluated node and the destination.

The routing algorithm in POLARIS is a point-to-point, time-dependent, and intermodal algorithm. The algorithm finds routes for the following modes:

\begin{itemize}
    \item Driving:
        \begin{itemize}
            \item Passenger cars
            \item Taxis and TNCs
        \end{itemize}
    \item Active:
        \begin{itemize}
            \item Walking
            \item Micromobility with personally owned bikes/scooters
            \item Shared micromobility with docked vehicles
            \item Shared micromobility with non-docked vehicles
        \end{itemize}
    \item Transit:
        \begin{itemize}
            \item Walk and transit
            \item Drive-to-transit (park-and-ride or kiss-and-ride)
            \item Drive-after-transit (Return transit trip to the parked car and drive to the destination)
            \item TNC-and-transit (TNC as a FMLM service or as a connector between transit legs)
            \item Micromobility-and-transit (Micromobility as a FMLM service or as a connector between transit legs)
        \end{itemize}
\end{itemize}

The point-to-point algorithm finds paths between two activity locations. Activity locations serve as nano-scale zones with a group of links associated with them. In other words, each activity location is projected onto a small set of walking, biking, and driving links. Based on the mode, the algorithm starts on a subset of these links. For instance, if the mode is passenger car, the algorithm starts on the subset of driving links, and transferring to another type of link throughout the algorithm is not possible. The algorithm considers a subset of destination driving links (associated with the destination activity location). The optimal starting link and the optimal destination link emerge from the algorithm. If the mode is walk and transit, the origin and destination subsets only include walking links. In the case of park-and-ride, the origin subset includes driving links only and the destination subset walking links. Transfer from a driving link to a walking link is only possible near a transit facility that provides parking, and the reverse is not possible. In case of drive-after-transit, the algorithm starts on walking links and finds a walk-to-transit path to where the car was parked. The second stage is a drive-only routing from the parking link to the destination activity location. If the mode is TNC-and-transit, both walking and driving links are allowed as the origin and destination link subsets. The reason for that is that the best path can be where TNC is the first leg, a bridge leg, the last leg, or any combination thereof, and the algorithm finds which combination and sequence of walking, driving, and transit is optimal.

\subsubsection{Simulation-Based Dynamic Traffic Assignment}\label[sec]{dta_details}

Capturing the circular relationship between routing decisions and network conditions is essential for every network model. Since network conditions change temporally, dynamic traffic assignment (DTA) models \cite{merchant1978model,merchant1978optimality} emerged as the appropriate tool to model the routing and movement of vehicles. This section omits the review of static assignment models due to space limitations. In simulation-based Dynamic Traffic Assignment (DTA), traffic is simulated rather than modeled analytically. Although this method is not as theoretically rigorous as analytical approaches \cite{peeta2001foundations}, it is better equipped to assess the space-time propagation of traffic flows, vehicle interactions, and the travel costs on both origin-destination routes and individual links \cite{lu2009equivalent}. The traffic assignment framework in POLARIS \cite{auld2016polaris} is another example of simulation-based DTA. Although the following list is not exhaustive, some well-known models are DYNASMART \cite{jayakrishnan1994evaluation,mahmassani2001dynamic}, DynaMIT \cite{ben2001network,ben2012dynamic}, DTALite \cite{zhouDTALite,Zhou_qu2017large}, MATSim \cite{balmer2009matsim,matsim_2016}, DYNAMEQ \cite{kettner2006examining}, and DynusT \cite{dynust}.

In every simulation-based transportation model, there is an iterative loop between finding shortest paths, assigning travelers to the shortest paths or keeping them on their existing paths, simulating the resulting traffic, and recalculating the shortest paths in the next iteration. Since there is no-closed form solution, i.e., no direct analytical expression of travel times as a function of link loads, simulation-based DTA tools rely on certain techniques when deciding whom to keep on their existing paths and whom to assign to newly-found shortest paths. As opposed to randomly selecting travelers in the method of successive averages (MAS) approach, gap-based approaches are introduced in \cite{sbayti2007efficient,lu2009dynamic, lu2009equivalent}. Gap-based approaches have proven to be very effective, since the travelers with a larger gap, i.e., larger difference in experienced and routed travel times in the previous iteration, are more likely to be assigned to the newly found shortest path. In their NUTRANS implementations, \citep{verbas2014dissertation, verbas2015trr_dta, verbas2016partb_dta} have implemented the same approach to the transit context.

POLARIS DTA framework builds on top of the gap-based approach using information mixing \cite{auld2019agent}. The approach is gap-based in two key aspects. First, similar to prior methods, the reassignment decision during the convergence process relies on the gap between the routed and experienced travel time from the previous iteration. Second, historical time-dependent and current traffic conditions are averaged into a single expected value for each agent using a weight derived from a modified two-parameter Weibull survival function. This weight is personalized based on the relative gap of the traveler from the previous iteration and the iteration number, which is a novel feature of POLARIS \cite{auld2019agent}.

\subsubsection{Agent-Based Transit Simulation}\label[sec]{simulation_details}

This subsection focuses on the movement of travelers and vehicles from a transit-centered perspective. The movement of vehicles on roadway traffic is not explained in this subsection, and interested readers are referred to \citep{deSouza2019mesoscopic, deSouza2024polarisLC}. The transit simulation module follows a similar process to the one in NUTRANS \cite{verbas2014dissertation, verbas2015hyperpath, verbas2015trr_dta, verbas2015integrated_freq_dta, verbas2016partb_dta}. The main extension on NUTRANS is the direct integration with the driving network, so that the travelers can perform fully intermodal movements, such as park-and-ride. Moreover, buses are able to run on the driving network in mixed-traffic conditions.

\begin{sidewaysfigure}[!htbp]
    \centering
    \includegraphics[width=1\textwidth,height=1\textheight,keepaspectratio]{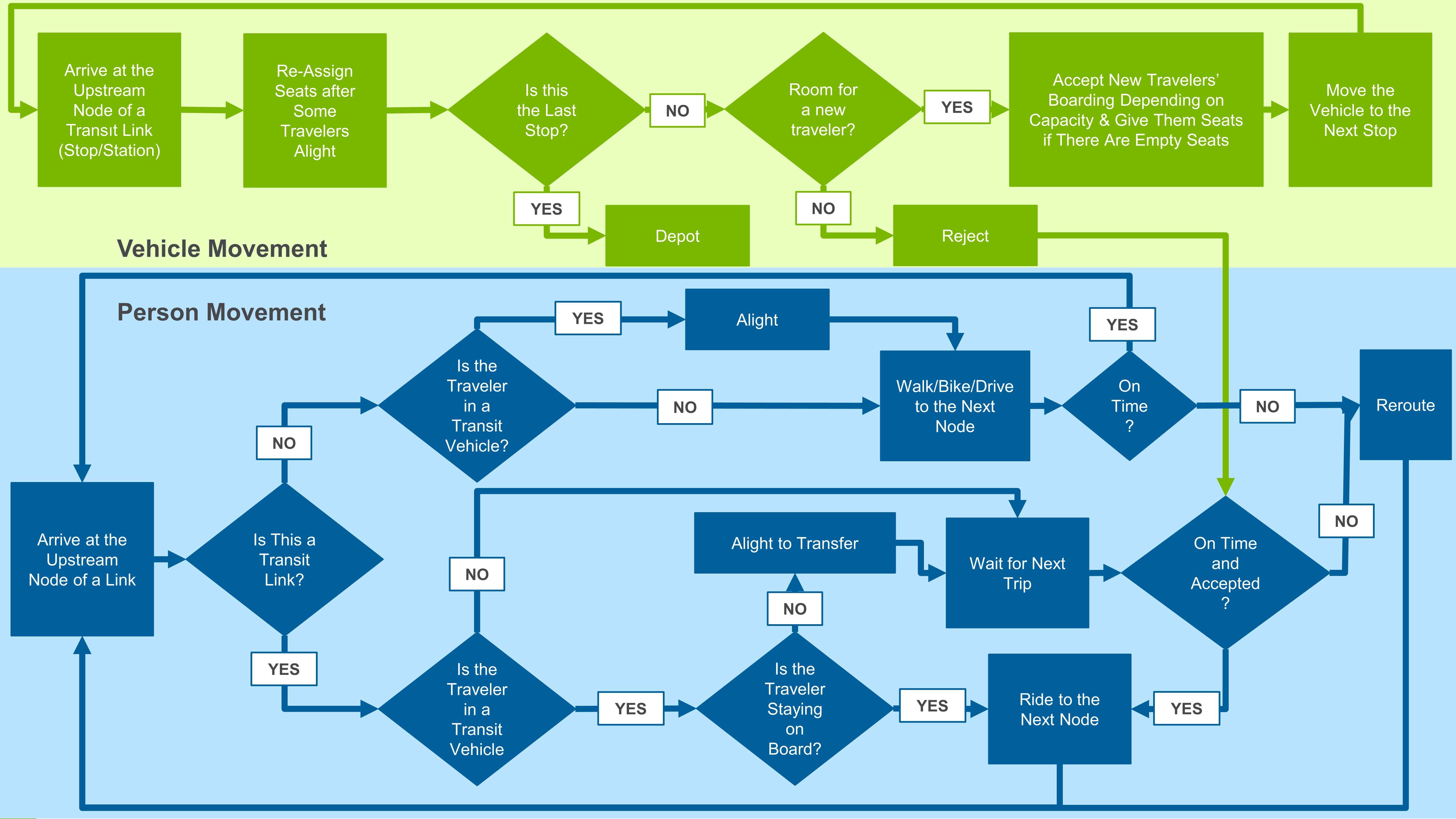}
    \caption{Transit simulation framework in POLARIS.}
    \label[fig]{transit_sim}
\end{sidewaysfigure}

Once a traveler is assigned to a route, they are moved in the network link-by-link. Similarly, transit vehicles move in the network based on either their pre-assigned (GTFS) schedule or based on mixed-traffic conditions. See \cref{transit_sim} for the flowchart of both person and transit vehicle submodules and their interaction.

For a given traveler, the link-by-link actions include:

\begin{itemize}
    \item Walking/biking/driving
    \item Waiting for a transit or TNC vehicle
    \item Boarding a transit or a TNC vehicle
    \item Traveling in a transit or TNC vehicle
        \begin{itemize}
            \item Standing
            \item Seating
        \end{itemize}
    \item Alighting
    \item Being rejected to board    
    \item Re-routing
    \item Picking (undocking) a micromobility vehicle
    \item Dropping (docking) a micromobility vehicle
\end{itemize}
\vspace{12pt}

For a given transit vehicle, the link-by-link actions include:

\begin{itemize}
    \item Moving
        \begin{itemize}
            \item In the transit network 
                \begin{itemize}
                    \item According to the GTFS schedule if train or mixed-traffic setting is off
                    \item Signalled by its twin in the traffic layer, if it is a bus and \textit{mixed-traffic} setting is on
                \end{itemize}
            \item In the driving network 
                \begin{itemize}
                    \item If it is a bus and \textit{mixed-traffic} setting is on, it signals its twin on the transit layer
                \end{itemize}
        \end{itemize}

    \item Accepting travelers to board
        \begin{itemize}
            \item Making them stand
            \item Making them take a seat
        \end{itemize}
    \item Letting travelers alight
    \item Rejecting travelers due to capacity
\end{itemize}
\vspace{12pt}

\begin{figure}[!h]
    \centering
    \includegraphics[width=1\linewidth]{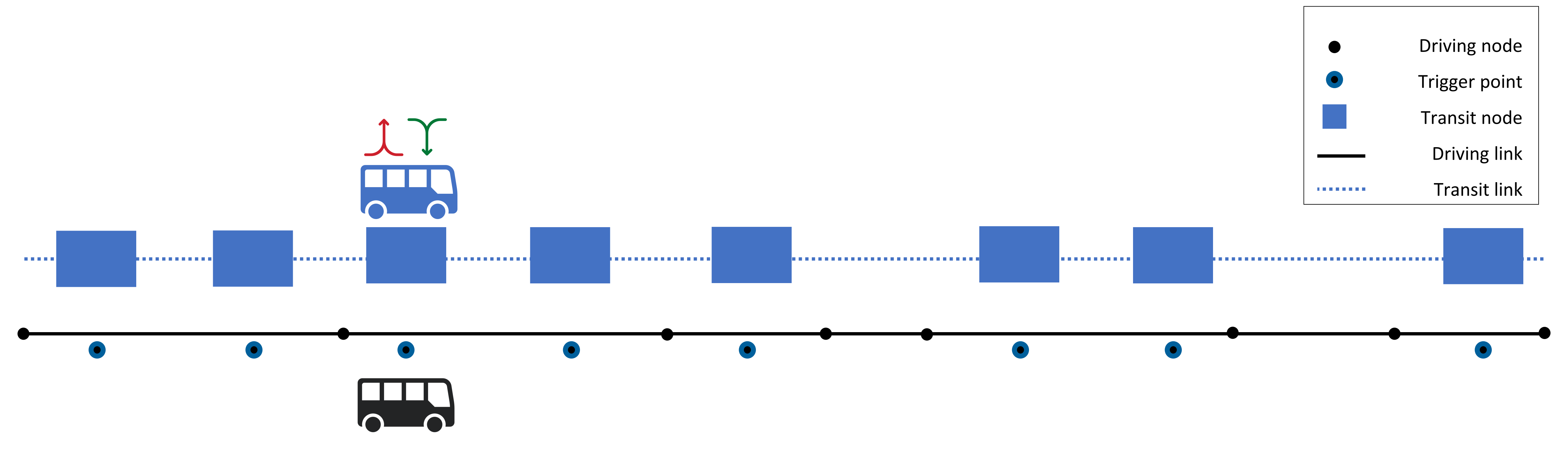}
    \caption{Running buses in mixed-traffic in POLARIS.}
    \label[fig]{buses}
\end{figure}

The parallel movement of buses in the transit and driving network layers is based on the \textit{Transit Pattern Mapping} table seen in \cref{db_schema}. The details are presented in \cref{buses}. Similar to \cref{network}, the transit nodes and links are in blue, whereas the driving nodes and links are in black. The \textit{trigger points} in black and blue correspond to each row in the \textit{Transit Pattern Mapping} table. When evaluating a transit link, the routing algorithm is only concerned with the transit link in the transit layer, although the travel time of the link now can be updated by the traffic conditions. There is always a bus agent (in blue) that dispatches from the first transit node of a transit trip at the GTFS scheduled time. If \textit{mixed-traffic} setting is off, it follows the GTFS schedule. Otherwise, as the black bus (the twin in the driving network) moves in the driving network, it crosses the trigger points one-by-one. Those trigger points correspond to the locations where the transit bus stop is projected on the driving link. Instead of arriving at the GTFS arrival time, the blue twin now appears on the next blue node when the black twin appears at its corresponding trigger point. The boarding and alighting movements happen on the blue twin, since transit passengers are moving along the transit network layer.

\section{Numerical Experiments}\label[sec]{exp_design}
In this section, we analyze the impact of various key levers on mode share and transit boardings considering potential supply and demand changes for the Chicago Metropolitan Region in 2040. With the exception of transit service improvement and rideshare/FMLM levers, these levers would not traditionally be associated with transit. However, in real-life, many transportation-related decisions have cascading effects, and POLARIS enables capturing those due to its multi-agent and multimodal framework. Here, we focus on the impact of these levers on transit. For an overall analysis, readers are referred to \citep{auld2024large}. The levers are summarized as follows:

\begin{itemize}
    \item \emph{Congestion pricing}: For all interstates, a fixed time-of-day toll charges are incurred based on expected delay per distance. 
    \item \emph{Vehicle electrification}: Three cases are considered based on different EV penetration rates for both light duty (LD) and heavy duty (HD) vehicles: (i) low - 15\% LD and 10\% HD, (ii) medium - 35\% LD and 25\% HD, (iii) high - 75\% LD and 50\% HD.
    \item \emph{E-commerce/on-demand delivery (ODD)}: Two cases are considered based on a year to year growth in e-commerce delivery demand: (i) a two percent growth in demand leading to 3.5 (e-commerce), 0.7 (ODD groceries), and 1.8 (ODD meals) deliveries per household per week and (ii) a six percent growth in demand corresponding to 9.8 (e-commerce), 1.5 (ODD groceries), and 3.5 (ODD meals) deliveries per household per week. Disabling this lever means using the former case while enabling it incorporates the latter.
    \item \emph{Transit service improvement}: Based on discussions with the Chicago Transit Authority (CTA) and the Regional Transportation Authority (RTA), 30\% speed increase on 39 CTA bus routes (resulting in nearly 30\% frequency increase), 40\% frequency increase for all Pace routes, and ensuring a 30-minute headway for all METRA routes between 6 AM and 10 PM - leading to 40\% overall frequency increase, are implemented. In addition, bus rapid transit (BRT) routes are introduced on Ashland and Western avenues for the CTA, and express (limited stop) bus routes on the 95th, Dempster, and Halsted streets are established for Pace.
    \item \emph{Rideshare/FMLM}: In lower travel demand density areas, TNC operators are assumed to provide a FMLM service from/to transit stops. The fare for such connections are subsidized by 50\% to improve access to congested areas.
    \item \emph{Connected infrastructure}: A local signal control scheme is implemented to improve the traffic flow. In this strategy, an early switch to green light is conducted if there are no vehicles approaching from opposing directions at cross intersections. This lever is henceforth refereed as \textit{Signals}.
\end{itemize}
\vspace{12pt}

\cref{boarding_regression}, \cref{boarding_regression_urban}, and \cref{boarding_regression_suburban} show the direct impact of statistically significant scenario levers on transit boardings, as well as the impact of significant lever interactions for the entire region, urban areas, and suburban areas, respectively. Through a regression analysis, it is found that incorporating congestion pricing, transit service improvement, FMLM subsidies, increased E-commerce/ODD, and vehicle electrification levers significantly impact transit boardings at the regional level. There are significant interactions between transit and the following: Pricing, signals, FMLM, and ODD. There is also significant interaction between FMLM and the following: Pricing, ODD. Looking closer at \cref{boarding_regression_urban} and \cref{boarding_regression_suburban}, we observe that most of the sensitivity in transit boardings is in the suburbs. The EV levers remain significant in the urban areas, whereas they lose their significance in the suburban areas. Contrarily, transit and pricing interaction, and transit and ODD interaction lose their significance in the urban areas, whereas they remain significant in the suburban areas.

\begin{table}[!h]
\footnotesize
\caption{Summary of regression analysis on transit boarding for the entire region}\label[tab]{boarding_regression}
\vspace{-0.8cm}
\begin{center}
\begin{tabular*}\textwidth{l@{\extracolsep{\fill}}ccccccc}
\toprule
  & Variable                 & Sensitivity (\%) & Coefficient  & Std. Err. & $t$       & $P>|t|$ & Significance \\
\cline{2-8}
  & Constant                 &                  & 2.291  & 0.004     & 630.102 & 0.000              & ***     \\
\cline{2-8}
\multirow{3}{*}{\rotatebox[origin=c]{90}{Supply}}       & Congestion Pricing       & 0.6              & 0.014  & 0.004     & 3.843   & 0.000              & ***     \\
  & Transit                  & 9.5              & 0.218  & 0.004     & 48.844  & 0.000              & ***     \\
  & FMLM                     & 3.8              & 0.088  & 0.004     & 20.968  & 0.000              & ***     \\
\cline{2-8}
\multirow{3}{*}{\rotatebox[origin=c]{90}{Demand}}       & ODD                      & -6.0             & -0.138 & 0.004     & -37.854 & 0.000              & ***     \\
  & EV (Medium)              & 0.4              & 0.009  & 0.003     & 3.402   & 0.001              & ***     \\
  & EV (High)                & 0.6              & 0.014  & 0.003     & 5.559   & 0.000              & ***     \\
\cline{2-8} 
\multirow{6}{*}{\rotatebox[origin=c]{90}{Interactions}} & Transit $\times$ Pricing & 0.4              & 0.010  & 0.004     & 2.451   & 0.014              & *       \\
  & Transit $\times$ Signals & 1.1              & 0.026  & 0.004     & 6.146   & 0.000              & ***     \\
  & Transit $\times$ FMLM    & 0.5              & 0.011  & 0.003     & 3.846   & 0.000              & ***     \\
  & Transit $\times$ ODD     & -0.6             & -0.013 & 0.004     & -3.104  & 0.002              & **      \\
  & FMLM $\times$ Pricing    & -0.7             & -0.015 & 0.004     & -3.684  & 0.000              & ***     \\
  & FMLM $\times$ ODD        & -1.2             & -0.028 & 0.004     & -6.645  & 0.000              & ***   \\ 
\bottomrule
\end{tabular*}
\end{center}
\vspace{-8pt}
Note: $N=765$, Adj. $R^2=0.959$, Significance codes: 0 `***', 0.001 `**', 0.01 `*', 0.05 `.', 0.1 `'
\end{table}

\begin{table}[!h]
\footnotesize
\caption{Summary of regression analysis on transit boarding for the urban areas}\label[tab]{boarding_regression_urban}
\vspace{-0.8cm}
\begin{center}
\begin{tabular*}\textwidth{l@{\extracolsep{\fill}}ccccccc}
\toprule
  & Variable                 & Sensitivity (\%) & Coefficient  & Std. Err. & $t$       & $P>|t|$ & Significance \\
\cline{2-8}
  & Constant                 &                  & 1.901  & 0.003     & 581.851 & 0.000              & ***     \\
\cline{2-8}
\multirow{3}{*}{\rotatebox[origin=c]{90}{Supply}}       & Congestion Pricing       & 0.3              & 0.006  & 0.003     & 1.811   & 0.071              & .     \\
  & Transit                  & 4.5              & 0.085  & 0.004     & 21.202  & 0.000              & ***     \\
  & FMLM                     & 3.7              & 0.070  & 0.004     & 18.539  & 0.000              & ***     \\
\cline{2-8}
\multirow{3}{*}{\rotatebox[origin=c]{90}{Demand}}       & ODD                      & -5.1             & -0.096 & 0.003     & -29.488 & 0.000              & ***     \\
  & EV (Medium)              & 0.4              & 0.008  & 0.002     & 3.524   & 0.000              & ***     \\
  & EV (High)                & 0.7            & 0.013  & 0.002     & 5.802   & 0.000              & ***     \\
\cline{2-8} 
\multirow{6}{*}{\rotatebox[origin=c]{90}{Interactions}} & Transit $\times$ Pricing & 0.2              & 0.003  & 0.004     & 0.817   & 0.414              &       \\
  & Transit $\times$ Signals & 0.4             & 0.007  & 0.003     & 2.776   & 0.006              & **     \\
  & Transit $\times$ FMLM    & -0.5              & -0.010  & 0.004     & -2.559   & 0.011              & *     \\
  & Transit $\times$ ODD     & -0.2             & -0.004 & 0.004     & -1.168  & 0.243             &       \\
  & FMLM $\times$ Pricing    & 0.9             & 0.018 & 0.004     & 4.690  & 0.000              & ***     \\
  & FMLM $\times$ ODD        & -1.2             & -0.023 & 0.004     & -6.080  & 0.000              & ***   \\ 
\bottomrule
\end{tabular*}
\end{center}
\vspace{-8pt}
Note: $N=765$, Adj. $R^2=0.896$, Significance codes: 0 `***', 0.001 `**', 0.01 `*', 0.05 `.', 0.1 `'
\end{table}

\begin{table}[!h]
\footnotesize
\caption{Summary of regression analysis on transit boarding for the suburban areas}\label[tab]{boarding_regression_suburban}
\vspace{-0.8cm}
\begin{center}
\begin{tabular*}\textwidth{l@{\extracolsep{\fill}}ccccccc}
\toprule
  & Variable                 & Sensitivity (\%) & Coefficient  & Std. Err. & $t$       & $P>|t|$ & Significance \\
\cline{2-8}
  & Constant                 &                  & 0.391  & 0.001     & 355.592 & 0.000              & ***     \\
\cline{2-8}
\multirow{3}{*}{\rotatebox[origin=c]{90}{Supply}}       & Congestion Pricing       &2.1              & 0.008  & 0.001     & 7.336   & 0.000              & ***     \\
  & Transit                  & 34.0              & 0.133  & 0.001     & 98.657  & 0.000              & ***     \\
  & FMLM                     & 4.6              & 0.018  & 0.001     & 14.281  & 0.000              & ***     \\
\cline{2-8}
\multirow{3}{*}{\rotatebox[origin=c]{90}{Demand}}       & ODD                      & -10.6             & -0.041 & 0.001     & -37.624 & 0.000              & ***     \\
  & EV (Medium)              & 0.2              & 0.001  & 0.001     & 0.782   & 0.435              &      \\
  & EV (High)                & 0.2            & 0.001 & 0.001     & 1.149   & 0.251              &      \\
\cline{2-8} 
\multirow{6}{*}{\rotatebox[origin=c]{90}{Interactions}} & Transit $\times$ Pricing & 1.8              & 0.007  & 0.001     & 5.684   & 0.000              &  ***     \\
  & Transit $\times$ Signals & 1.0             & 0.004  & 0.001     & 4.478   & 0.000             & ***     \\
  & Transit $\times$ FMLM    & -0.9              & -0.003  & 0.001     & -2.666   & 0.008              & **     \\
  & Transit $\times$ ODD     & -2.8             & -0.011 & 0.001     & -8.726  & 0.000            &      *** \\
  & FMLM $\times$ Pricing    & 2.1             & 0.008 & 0.001     & 6.399  & 0.000              & ***     \\
  & FMLM $\times$ ODD        & -1.3             & -0.005 & 0.001     & -3.918  & 0.000              & ***   \\ 
\bottomrule
\end{tabular*}
\end{center}
\vspace{-8pt}
Note: $N=765$, Adj. $R^2=0.959$, Significance codes: 0 `***', 0.001 `**', 0.01 `*', 0.05 `.', 0.1 `'
\end{table}

\begin{figure*}[h]
\centering
\subfloat[mode share.\label{fig:congestion_pricing_mode_share}]{%
\includegraphics*[width=0.48\textwidth,height=\textheight,keepaspectratio]{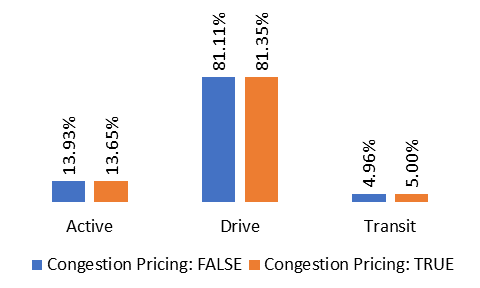}}
\quad
\subfloat[transit boardings.\label{fig:congestion_pricing_boarding}]{%
\includegraphics*[width=0.48\textwidth,height=\textheight,keepaspectratio]{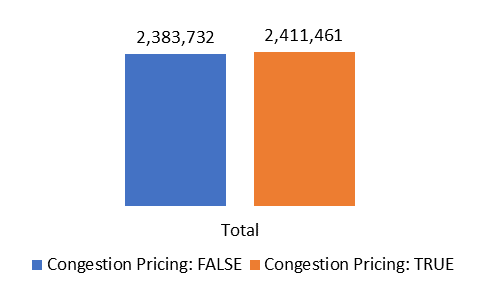}}
\caption{Impact of congestion pricing on.} \label[fig]{congestion_pricing}
\end{figure*}

\cref{congestion_pricing} shows the impact of enabling congestion pricing on mode share of transit and transit boardings. As seen in \cref{fig:congestion_pricing_mode_share}, transit mode share increases by 0.75\%, while transit boardings increase by 1.16\% (\cref{fig:congestion_pricing_boarding}). While the impact is numerically small, it is statistically significant.

\begin{figure*}[h]
\centering
\subfloat[mode share.\label{fig:EV_penetration_mode_share}]{%
\includegraphics*[width=0.48\textwidth,height=\textheight,keepaspectratio]{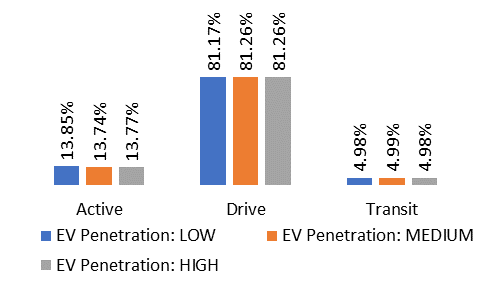}}
\quad
\subfloat[transit boardings.\label{fig:EV_penetration_boarding}]{%
\includegraphics*[width=0.48\textwidth,height=\textheight,keepaspectratio]{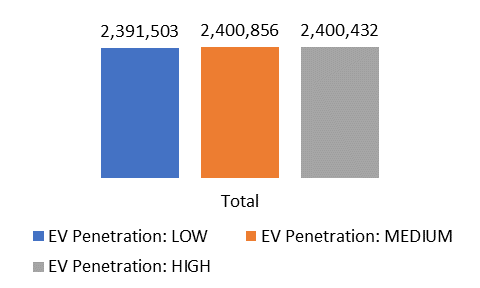}}
\caption{Impact of vehicle electrification on.} \label[fig]{EV_penetration}
\end{figure*}

\cref{EV_penetration} depicts the impact of personal EV stock share on mode share (\cref{fig:EV_penetration_mode_share}) and boardings (\cref{fig:EV_penetration_boarding}). Medium and high EV stock share  significantly impacts transit boardings in urban areas and the entire region as supported by \cref{boarding_regression_urban} and \cref{boarding_regression}, respectively; however, the impact in suburban areas is insignificant (see \cref{boarding_regression_suburban}).

\begin{figure*}[h]
\centering
\subfloat[mode share.\label{fig:E_commerce_mode_share}]{%
\includegraphics*[width=0.48\textwidth,height=\textheight,keepaspectratio]{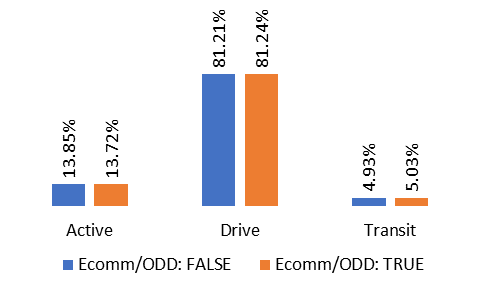}}
\quad
\subfloat[transit boardings.\label{fig:E_commerce_boarding}]{%
\includegraphics*[width=0.48\textwidth,height=\textheight,keepaspectratio]{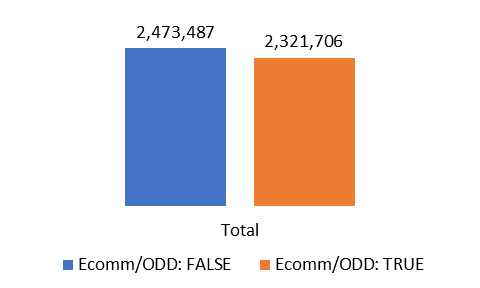}}\\
\subfloat[number of trips.\label{fig:E_commerce_num_trips}]{%
\includegraphics*[width=0.48\textwidth,height=\textheight,keepaspectratio]{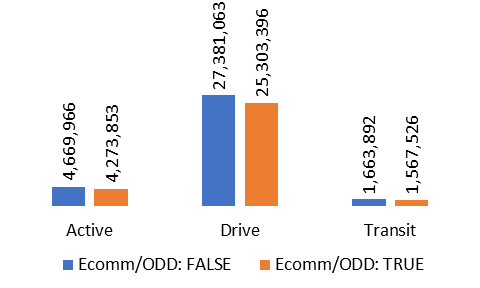}}
\caption{Impact of E-commerce/ODD on.} \label[fig]{E_commerce}
\end{figure*}

The impact of e-commerce delivery/ODD is demonstrated in \cref{E_commerce}. As seen in \cref{fig:E_commerce_num_trips}, the increase in e-commerce deliveries reduces the total number of trips across all modes. The total reduction is around 2.57M (7.62\%).  As a result, transit boardings decrease, as well, by around 152K (6.14\%) (see \cref{fig:E_commerce_boarding}). However, since the relative reduction in driving trips are higher, the mode share of transit increases by 1.98\% as seen in \cref{fig:E_commerce_mode_share}.

\begin{figure*}[h]
\centering
\subfloat[mode share.\label{fig:Transit_improvement_mode_share}]{%
\includegraphics*[width=0.48\textwidth,height=\textheight,keepaspectratio]{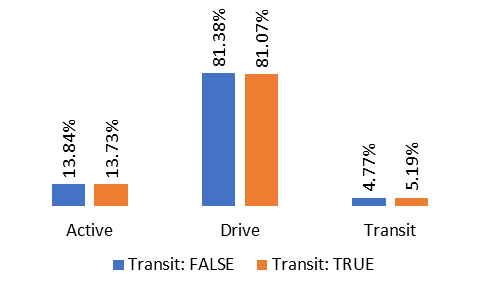}}
\quad
\subfloat[transit boardings.\label{fig:Transit_improvement_boarding}]{%
\includegraphics*[width=0.48\textwidth,height=\textheight,keepaspectratio]{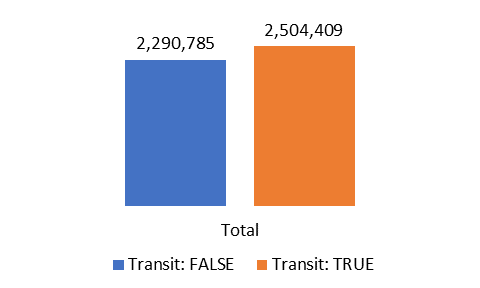}}
\caption{Impact of transit service improvement on.} \label[fig]{Transit_improvement}
\end{figure*}

\cref{Transit_improvement} illustrates the contribution of transit service improvement. The transit mode share and boardings increase by 8.86\% (around 145K) and 9.33\% (around 214K), respectively (see \cref{fig:Transit_improvement_mode_share} and \cref{fig:Transit_improvement_boarding}). The reason behind the higher increase in boardings is due to a higher number of transfers. Improved transit service with higher frequencies, more coverage, and higher speeds facilitates easier transfers.

\begin{figure*}[h]
\centering
\subfloat[mode share.\label{fig:Ridesharing_mode_share}]{%
\includegraphics*[width=0.48\textwidth,height=\textheight,keepaspectratio]{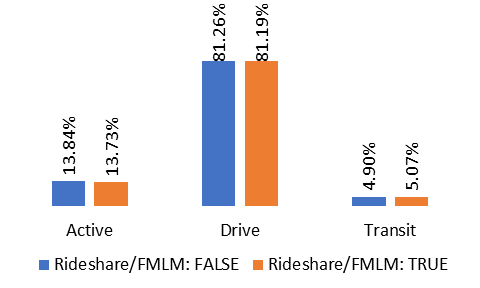}}
\quad
\subfloat[transit boardings.\label{fig:Ridesharing_boarding}]{%
\includegraphics*[width=0.48\textwidth,height=\textheight,keepaspectratio]{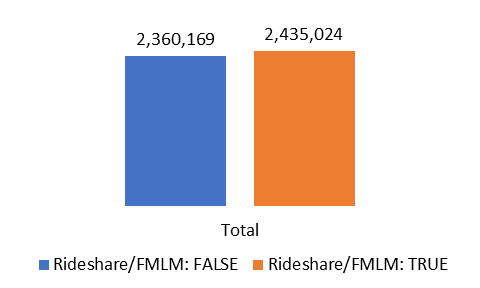}}
\caption{Impact of rideshare/FMLM on.} \label[fig]{Ridesharing}
\end{figure*}

In \cref{Ridesharing}, the impact of rideshare/FMLM is presented. In this scenario, transit mode share and boardings increase by 3.58\% (around 57K) (\cref{fig:Ridesharing_mode_share}) and 3.13\% (around 75K) (\cref{fig:Ridesharing_boarding}), respectively.

\begin{figure*}[!h]
\centering
\subfloat[mode share.\label{fig:Ridesharing_transit_mode_share}]{%
\includegraphics*[width=0.48\textwidth,height=\textheight,keepaspectratio]{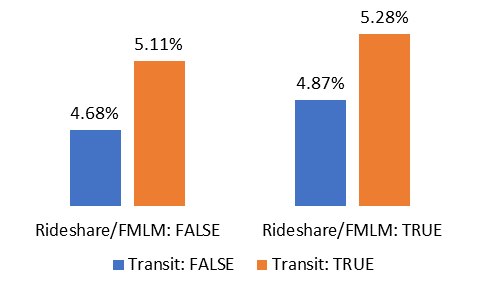}}
\quad
\subfloat[transit boardings.\label{fig:Ridesharing_transit_boarding}]{%
\includegraphics*[width=0.48\textwidth,height=\textheight,keepaspectratio]{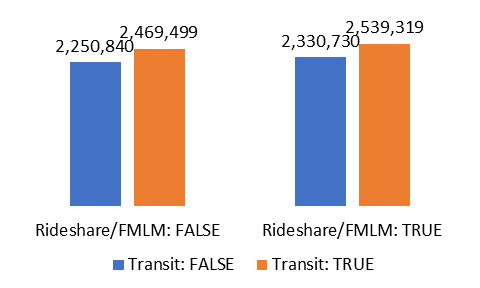}}
\caption{Impact of rideshare/FMLM combined with transit service improvement on.} \label[fig]{Ridesharing_transit}
\end{figure*}

\cref{Ridesharing_transit} presents the joint impact of the rideshare/FMLM subsidization and the transit service improvement. Studying \cref{fig:Ridesharing_transit_mode_share} closer, it is seen that:

\vspace{12pt}
\begin{itemize}
    \item Turning Rideshare/FMLM lever true increases transit mode share by:
        \begin{itemize}
            \item 3.98\% when transit lever is false
            \item 3.22\% when transit lever is true
        \end{itemize}
    \item Turning Transit lever true increases transit mode share by:
        \begin{itemize}
            \item 9.27\% when Rideshare/FMLM lever is false
            \item 8.47\% when Rideshare/FMLM lever is true
        \end{itemize}
    \item Turning both levers from false to true increases transit mode share by 12.78\%
\end{itemize}
\vspace{12pt}

Studying \cref{fig:Ridesharing_boarding} closer, it is seen that:

\vspace{12pt}
\begin{itemize}
    \item Turning Rideshare/FMLM lever true increases transit boardings share by:
        \begin{itemize}
            \item 3.55\% when transit lever is false
            \item 2.83\% when transit lever is true
        \end{itemize}
    \item Turning Transit lever true increases transit boardings by:
        \begin{itemize}
            \item 9.71\% when Rideshare/FMLM lever is false
            \item 8.95\% when Rideshare/FMLM lever is true
        \end{itemize}
    \item Turning both levers from false to true increases transit boardings share by 12.82\%
\end{itemize}
\vspace{12pt}

\section{Conclusion}\label[sec]{conclusion}
Public transit is a core component of transportation simulation tools. In POLARIS, the transit network is simulated alongside other travel modes such as automobiles, bikes, trucks, and walking. POLARIS is a high-performance, computationally efficient agent-based simulation tool that incorporates advanced travel and freight demand modeling, dynamic traffic and transit assignment, and transportation simulation within a unified modeling platform. This study concentrates on the transit routing, assignment, and simulation components of POLARIS, presenting its structural design and explaining key terminologies and concepts used. The model fully integrates upstream decision-making processes of travelers, such as activity generation, location and timing choices, and mode choice (focusing on trips that include transit). These decisions are followed by routing and assignment decisions and the movement of travelers and vehicles within the multimodal network. This integrated framework allows for comprehensive modeling of interactions between all agents, including travelers, vehicles, and transportation service providers.

In this study, we provided a literature review on existing transportation system modeling tools, explained the details on transit modeling framework embedded in POLARIS while pointing out to key theoretical concepts, and summarized the findings of a large-scale analysis focusing on multiple supply and demand lever combinations. In the numerical experiments conducted, we observed policy interventions such as congestion pricing, transit service improvement, first-mile-last-mile subsidies for transportation network company trips, increased e-commerce/on-demand deliveries, and vehicle electrification have significant impacts on transit ridership. Using some of these levers together have also some synergistic or cancelling effects. The case study demonstrates the importance of modelling transit in a much larger context, which is the full integration of multimodal network simulation and the decision-making of all agents.

\section*{Acknowledgements}
This material is based on work supported by the U.S. Department of Energy, Office of Science, under contract number DE-AC02-06CH11357. This report and the work described were sponsored by the U.S. Department of Energy (DOE) Vehicle Technologies Office (VTO) under the Transportation Systems and Mobility Tools Core Maintenance/Pathways to Net-Zero Regional Mobility, an initiative of the Energy Efficient Mobility Systems (EEMS) Program. Erin Boyd, a DOE Office of Energy Efficiency and Renewable Energy (EERE) manager, played an important role in establishing the project concept, advancing implementation, and providing guidance.

\section*{Author Contributions}
The authors confirm their contribution to the paper as follows: agent-based network simulation development: all authors; multimodal network representation and generation: Pedro Veiga de Camargo, Omer Verbas, Taner Cokyasar; study conception: all authors; data collection: all authors; designing simulation input and running simulations: all authors; analysis and interpretation of results: all authors; draft manuscript preparation: Omer Verbas, Taner Cokyasar, Pedro Veiga de Camargo. All authors reviewed the results and approved the final version of the manuscript.

\clearpage
\bibliographystyle{trb} 
\bibliography{trb_template}

\vfill
\framebox{\parbox{.90\linewidth}{\scriptsize The submitted manuscript has been created by
        UChicago Argonne, LLC, Operator of Argonne National Laboratory (``Argonne'').
        Argonne, a U.S.\ Department of Energy Office of Science laboratory, is operated
        under Contract No.\ DE-AC02-06CH11357.  The U.S.\ Government retains for itself,
        and others acting on its behalf, a paid-up nonexclusive, irrevocable worldwide
        license in said article to reproduce, prepare derivative works, distribute
        copies to the public, and perform publicly and display publicly, by or on
        behalf of the Government.  The Department of Energy will provide public access
        to these results of federally sponsored research in accordance with the DOE
        Public Access Plan \url{http://energy.gov/downloads/doe-public-access-plan}.}}
\end{document}